\documentclass[12pt]{article}
\usepackage{amssymb}
\newcommand{\be}{\begin{eqnarray}}
\newcommand{\ee}{\end{eqnarray}}

\renewcommand{\d}{{\rm d}}
\newcommand{\dl}{\delta}
\newcommand{\D}{{\rm D}}
\newcommand{\Dl}{{\cal D}}
\title{\bf On generalized Lotka-Volterra lattices}
\author{A. Dimakis$^1$ and F. M\"uller-Hoissen$^2$}
\date{  }
\begin{document}
\maketitle
\begin{center}
 $^1$ Dept. Financial and Management Engineering,
 Aegean University, GR-82100 Chios
 \vskip.1cm
 $^2$ Max-Planck-Institut f\"ur Str\"omungsforschung,
 Bunsenstrasse 10, D-37073 G\"ottingen
\end{center}

\vskip.1cm

\begin{abstract}
Generalized matrix Lotka-Volterra lattice equations are obtained in a
systematic way from a ``master equation" possessing a bicomplex
formulation.
\end{abstract}

\section{Introduction}
A {\em bicomplex} is an ${\mathbb{N}}_0$-graded linear space
(over ${\mathbb{R}}$) ${\cal M} = \bigoplus_{s \geq 0} {\cal M}^s$
together with two linear maps
$\Dl , \D \, : \, {\cal M}^s \rightarrow {\cal M}^{s+1}$
satisfying $(\Dl - \lambda \, \D)^2 = 0$ for all $\lambda \in \mathbb{R}$.
If this holds as a consequence of a certain (partial
differential or difference) equation, we speak of a
\emph{bicomplex formulation} for this equation \cite{DMH-bc,DMH-BT}.
It may be regarded as a special case of a parameter-dependent
zero curvature condition.
In this work we derive generalized Lotka-Volterra (LV) lattices
(see \cite{Nari82,Itoh87,Bogo,Hu+Bull97,Suri99}, for example) from a
``master equation" with a bicomplex formulation. Some consequences of
the latter (conservation laws \cite{DMH-bc}, B\"acklund
transformations \cite{DMH-BT}) are briefly discussed.
We choose the bicomplex space as ${\cal M} = {\cal M}^0 \otimes \Lambda$
where $\Lambda = \bigoplus_{s=0}^2 \Lambda^s$
is the exterior algebra of a $2$-dimensional vector space
with a basis $\xi, \eta$ of $\Lambda^1$.
It is sufficient to define bicomplex maps on ${\cal M}^0$,
since by linearity and action on the coefficients of
monomials in $\xi, \eta$ they extend to the whole of
${\cal M}$ (cf \cite{DMH-bc,DMH-BT}).
On ${\cal M}^0$ we define
\be
  \dl z = (z - S^{q-p} z) \, \xi + \dot{z} \, \eta \, , \qquad
  \d  z = (S^q z - z) \, \xi + (S^p z - z) \, \eta  \label{triv_bic}
\ee
where $\dot{z} = dz/dt$, $S$ is an invertible linear operator on
${\cal M}^0$ with $\dot{S} = 0$, and $p,q \in \mathbb{Z}$.
The bicomplex equations for $({\cal M},\dl,\d)$ are then identically
satisfied. In the following, the maps $\dl$ and $\d$ will be
``dressed", resulting in nontrivial bicomplexes.

\section{A class of generalized Lotka-Volterra equations and its properties}

\subsection{A bicomplex for generalized LV equations}
Let us replace $\d$ by a map $\D$ which acts as follows on ${\cal M}^0$,
\be
  \D z = (Q \, S^q z - z) \, \xi + (P \, S^p z - z) \, \eta
         \label{d-dressed}
\ee
where $Q$ and $P$ are linear operators on ${\cal M}^0$. Then $({\cal M},
\dl, \D)$ is a bicomplex iff
\be
  Q S^q P S^p = P S^p Q S^q \, , \qquad
  \dot{Q} S^q = P S^q - S^{q-p} P S^p \, .   \label{bcQP}
\ee
Choosing ${\cal M}^0$ as the set of $\mathbb{C}^m$-valued
functions (on an infinite lattice) $z_n(t)$, $n \in \mathbb{Z}$,
and $(S z)_n = z_{n+1}$, $(Q z)_n = Q_n \, z_n$, $(P z)_n = P_n \, z_n$,
(\ref{bcQP}) becomes
\be
  Q_n P_{n+q} = P_n Q_{n+p} \, , \qquad
  \dot{Q}_n = P_n -P_{n+q-p} \; .
\ee
The ansatz
\be
  Q_n = g_n^{-1} g_{n+q} \, , \qquad
  P_n = g_n^{-1} g_{n+p}     \label{PQ2g}
\ee
with invertible matrices $g_n$ now solves the first equation and turns
the second into
\be
   (g_n^{-1} g_{n+q}) \, \mathbf{\dot{ }} \,
 = g_n^{-1} \, g_{n+p} - g_{n+q-p}^{-1} \, g_{n+q} \, .  \label{gbo}
\ee
In the following we exclude the trivial cases where $p=0$ or $q=0$.
If $p$ and $q$ are not coprime, let $s$ denote their greatest common
divisor, so that $p = s p'$ and $q = s q'$ where $p',q'$ are coprime.
The above equation then decomposes into independent equivalent equations
on $s$ sublattices. Therefore we restrict $q$ and $p$ to be coprime.
Since $\{ p,q,n, g_n \} \mapsto \{ -p,-q,-n, g_{-n} \}$
is a symmetry of (\ref{gbo}), it is sufficient
to consider $q > 0$.
Generalized LV equations are now obtained as follows. Introducing
\be
    V_n := g_n^{-1} \, g_{n+1}        \label{defV}
\ee
equation (\ref{gbo}) takes the form
\be
   (V_n \cdots V_{n+q-1}) \, \mathbf{\dot{ }} \,
 = \left\{ \begin{array}{l@{\quad\mbox{if }}l}
    V_n \cdots V_{n+p-1} - V_{n+q-p} \cdots V_{n+q-1} & p>0 \\
    V_{n-1}^{-1} \cdots V_{n+p}^{-1} - V_{n+q-p-1}^{-1} \cdots V_{n+q}^{-1}
    & p<0 \end{array} \right.               \label{LV}
\ee
In the case where $V_n$ are scalars, we write $a_n$ instead of $V_n$
and the last two equations turn out to be the ``extended Lotka-Volterra
equations" studied in \cite{Nari82,Hu+Bull97}.
For $q=1$, they reduce to
$\dot{a}_n = a_n \, (a_{n+1} \cdots a_{n+p-1} - a_{n-p+1} \cdots a_{n-1})$
and
$\dot{a}_n = a_{n-1}^{-1} \cdots a_{n+p}^{-1} - a_{n+1}^{-1}
 \cdots a_{n-p}^{-1}$,
respectively, which have been explored in particular in \cite{Bogo,Suri99}.

\noindent
{\em Remark.} In case of a finite open lattice of $N$ points, choose
$S = \sum_{i=1}^{N-1} E_{i,i+1}$ where $E_{i,j}$ are
the elementary matrices with a $1$ at the $i$th row and $j$th column and
otherwise zeros, and define $S^{-1}$ as the transpose of $S$.
Since $[S,S^T] \neq 0$, (\ref{triv_bic}) does not yield a bicomplex.
But after dressing, a bicomplex $({\cal M},\dl,\D)$ for the finite open
LV equation is obtained. For a periodic lattice choose
$S = \sum_{i=1}^{N-1} E_{i,i+1} + E_{N,1}$.
\hfill $\blacksquare$

\subsection{The bicomplex linear system and conservation laws}
The \emph{bicomplex linear system} \cite{DMH-bc} for
the bicomplex introduced above is
$\dl \chi = \lambda \, \D \chi$ with an $m \times m$ matrix field $\chi$.
The integrability condition $\dl \D \chi = 0$ leads to
\be
   (Q_n \, \chi_{n+q} - \chi_n) \, \mathbf{\dot{ }} \,
 = P_n \, \chi_{n+p} - \chi_n
   - (P_{n+q-p} \, \chi_{n+q} - \chi_{n+q-p})
\ee
which has the form of a conservation law. As a consequence,
\be
 {\cal Q}_i = \lambda \sum_{k=-\infty}^{+\infty} (Q_{i+k(q-p)} \, \chi_{i+q+k(q-p)}
  - \chi_{i+k(q-p)})
\ee
where $i=0,1,\ldots,|q-p|-1$, are conserved charges, i.e. $\dot{\cal Q}_i = 0$.
The linear system reads
$\chi_n - \chi_{n+q-p} = \lambda \, (Q_n \, \chi_{n+q} - \chi_n)$,
$\dot{\chi}_n = \lambda \, (P_n \, \chi_{n+p} - \chi_n)$.
These equations can be solved iteratively with a power series expansion
$\chi = \sum_{r=0}^\infty \lambda^r \chi^{(r)}$. Starting with $\chi^{(0)}=I$,
the unit matrix, we find the first set of conserved quantities
\be
  {\cal Q}_i^{(1)} = \sum_{k=-\infty}^{+\infty} (Q_{i+k(q-p)} -I) \, .
\ee
Solving the linear system for $\chi^{(1)}$ leads to
$\chi^{(1)}_n - \chi^{(1)}_{n+q-p} = Q_n - I$ and thus
\be
   \chi^{(1)}_n = \sum_{j=1}^\infty ( I - Q_{n-j(q-p)} )
\ee
with the help of which one obtains the second set of conserved charges,
\be
    {\cal Q}^{(2)}_i = \sum_{k=-\infty}^{+\infty} \Big( Q_{i+k(q-p)}
      \sum_{j=1}^{\infty} (I - Q_{i+q+(k-j)(q-p)})
      - \sum_{j=1}^{\infty} (I - Q_{i+(k-j)(q-p)}) \Big)
\ee
and so forth. Expressing $Q_n$ in terms of $V_n$ via (\ref{PQ2g}) and
(\ref{defV}), conserved charges are obtained for the LV equations
(\ref{LV}).

\subsection{B\"acklund transformations}
A ``Darboux-B\"acklund transformation'' (DBT) of a bicomplex leads to
a B\"acklund transformation (BT) of the associated equation \cite{DMH-BT}.
In order to determine a (primary) auto-DBT
of the bicomplex associated with (\ref{gbo}), one has to find an
operator $R_{21}$ on ${\cal M}^0$ which solves the equations
$[\dl,R_{21}] = \D_2 - \D_1$ and $\D_2 \, R_{21} = R_{21} \, \D_1$
where $\D_1, \D_2$ are the bicomplex map $\D$ depending on solutions
$g_1$ and $g_2$ of (\ref{gbo}), respectively.
With $R_{21} = r \, S^p$, where $r$ is a function, this leads to
\be
  r_n - r_{n+q-p} &=& Q_{2,n} - Q_{1,n} \, , \qquad
  \dot{r}_n = P_{2,n} - P_{1,n}  \label{b1,2} \\
  Q_{2,n} \, r_{n+q} &=& r_n \, Q_{1,n+p} \, , \qquad
  P_{2,n} \, r_{n+p} = r_n \, P_{1,n+p} \, . \label{b3,4}
\ee
Using (\ref{PQ2g}), (\ref{b3,4}) becomes $A_{n+q} = A_n = A_{n+p}$
where $A_n = g_{2,n} \, r_n \, g_{1,n+p}^{-1}$.
Since $p,q$ are relatively prime, according to the Euclidean
algorithm there are integers $k,l$ such that
$k q - l p = 1$. Hence $A_{n+1} = A_{n+1 + l p} = A_{n+k q} = A_n$
and thus
\be
  r_n = g_{2,n}^{-1} \, A(t) \, g_{1,n+p}
\ee
with an arbitrary matrix $A$ depending on $t$ only.
(\ref{b1,2}) now produces a BT:
\be
 g_{2,n}^{-1} \, A \, g_{1,n+p} - g_{2,n+q-p}^{-1} \, A \, g_{1,n+q}
 &=& g_{2,n}^{-1} \, g_{2,n+q} - g_{1,n}^{-1} \, g_{1,n+q} \\
 (g_{2,n}^{-1} \, A \, g_{1,n+p}) \, \mathbf{\dot{ }} \,
 &=& g_{2,n}^{-1} \, g_{2,n+p} - g_{1,n}^{-1} \, g_{1,n+p}  \, .
\ee
In the case under consideration, the ``permutability theorem" takes
the form $R_{31}+R_{10}=R_{32}+R_{20}$ (see \cite{DMH-BT} for details).
This leads to the superposition formula
\be
 g_{3,n} = (A_1 \, g_{2,n+p} - A_2 \, g_{1,n+p}) \, g_{0,n+p}^{-1} \,
 (g_{1,n}^{-1} \, A_1 - g_{2,n}^{-1} \, A_2)^{-1} \, .  \label{pec}
\ee
In the scalar case, the above BT can be transformed into Hirota's
bilinear form \cite{DMH-BT}. BTs of the scalar generalized LV equations
have been derived previously in \cite{Hu+Bull97}.

\subsection{Another class of generalized LV equations}
Instead of $\d$, now we dress the operator $\dl$ of the trivial bicomplex
$({\cal M}, \d, \dl)$:
\be
 \Dl z = (z - L S^{q-p}z) \, \xi + (\dot{z} - M z) \, \eta
\ee
with linear operators $L$ and $M$.
The bicomplex conditions for $({\cal M},\Dl,\d)$ are
\be
 S^q M - M S^q = S^p L S^{q-p} - L S^q \, , \qquad
 \dot{L} S^{q-p} = M L S^{q-p} - L S^{q-p} M \, .
\ee
Choosing $(S z)_n = z_{n+1}$, $(L z)_n = L_n z_n$ and $(M z)_n = M_n z_n$,
this takes the form
\be
  M_{n+q} - M_n = L_{n+p} - L_n \, , \quad
  \dot{L}_n = M_n L_n - L_n M_{n+q-p} \, .  \label{LVbc2}
\ee

\noindent
{\em Remark.}
$L_n = g_{n+p} \, g_{n+q}^{-1}$ and $M_n = \dot{g}_{n+p} \,  g_{n+p}^{-1}$
solve (\ref{LVbc2}b) and (\ref{LVbc2}a) becomes (\ref{gbo}), so that
the bicomplex conditions for $({\cal M},\dl,\D)$ and $({\cal M},\Dl,\d)$
coincide. Then $\Dl = g \, \dl \, g^{-1}$ and $\d = g \, \D \, g^{-1}$,
so that the two bicomplexes are gauge equivalent.
\hfill $\blacksquare$
\vskip.1cm

The first equation of (\ref{LVbc2}) can be solved by setting
\be
  M_n = G_{n+p} - G_n \, , \qquad
  L_n = G_{n+q} - G_n
\ee
with matrices $G_n(t)$. The second equation then reads
\be
    (G_{n+q} - G_n) \, \mathbf{\dot{ }} \,
  = (G_{n+p} - G_n) (G_{n+q} - G_n)
    - (G_{n+q} - G_n) (G_{n+q} - G_{n+q-p}) \, . \label{G-eq}
\ee
Again we should restrict $p,q$ to be coprime and different
from zero. Since the map $\{ p,q,n,G_n \} \mapsto \{ -p,-q,-n,G_{-n} \}$
is a symmetry of (\ref{G-eq}), it is sufficient to consider $q>0$.
Moreover, the substitution $p = q-p'$ turns
(\ref{G-eq}) with $p<0$ into the same equation where $G_n$ is replaced
by $G^T_n$ and $p$ replaced by $p'$. Since $p'>q$, we thus only
need to consider $p,q>0$.
Introducing $W_n := G_{n+1}-G_n$ we obtain
\be
   {d \over dt} \sum_{i=0}^{q-1} W_{n+i}
 = \Big( \sum_{j=0}^{p-1} W_{n+j} \Big)
  \sum_{i=0}^{q-1} W_{n+i} - \Big( \sum_{i=0}^{q-1} W_{n+i} \Big)
  \sum_{j=1-q}^{p-q} W_{n-j}  \; .  \label{LV3}
\ee
The scalar case with $q=1$ and $p>1$ has been studied
by various authors \cite{Itoh87,Bogo,Suri99}.

\section{Discrete time generalized Lotka-Volterra equations}
\noindent
{\bf A.} In order to obtain discrete time counterparts
(see \cite{Suri99,Suri96})
of the generalized LV equations in section 2, we simply have to replace
the trivial bicomplex $({\cal M}, \dl, \d)$ introduced above
by a time-discretized  version:
\be
  \dl z = (z - T S^{q-p} z) \, \xi + h^{-1} \, (T z - z) \, \eta
  \, , \qquad
  \d  z = (S^q z - z) \, \xi + (S^p z - z) \, \eta  \label{tdbc}
\ee
where ${\cal M}^0$ is now chosen as the set of $\mathbb{C}^m$-valued
functions $z_n(t)$ depending on two discrete variables
$n \in \mathbb{Z}, \, t \in h \mathbb{Z}$ with
$h \in \mathbb{R} \setminus \{ 0 \}$.
The operators $S$ and $T$ act on ${\cal M}^0$ according to
$(Sz)_n(t) = z_{n+1}(t)$ and $(Tz)_n(t) = z_n(t+h) =: \tilde{z}_n(t)$.
Next we dress $\d$ as in (\ref{d-dressed}) with
$(Qz)_n = Q_n \, z_n$ and $(Pz)_n = P_n \, z_n$.
The resulting bicomplex equations are
\be
   Q_n \, P_{n+q} = P_n \, Q_{n+p} \, , \qquad
   \tilde{Q}_n - Q_n = h \, (P_n - \tilde{P}_{n+q-p}) \; .
\ee
The ansatz (\ref{PQ2g})
solves the first equation and turns the second into
\be
   \tilde{g}_n^{-1} \, \tilde{g}_{n+q} - g_n^{-1} \, g_{n+q}
 = h \, ( g_n^{-1} \, g_{n+p} - \tilde{g}_{n+q-p}^{-1} \, \tilde{g}_{n+q}) \, .
    \label{gdbo}
\ee
Again, we may restrict to coprime $p,q$ and $q > 0$.
Using (\ref{defV}), we obtain
\be
     (I + h \, \tilde{V}_{n+q-p} \cdots \tilde{V}_{n-1}) \,
     \tilde{V}_n \cdots \tilde{V}_{n+q-1}
 = V_n \cdots V_{n+q-1} \, (I + h \, V_{n+q} \cdots V_{n+p-1})
        \label{dtLV1}
\ee
if $p > q$. In the scalar case with $q=1$ we recover equation
(3.5) in \cite{Suri96}:
\be
 \tilde{a}_n \, \Big(1 + h \prod_{i=1}^{p-1} \tilde{a}_{n-i} \Big)
 = a_n \, \Big(1 + h \prod_{i=1}^{p-1} a_{n+i} \Big) \, .
\ee
If $0 < p < q$, (\ref{gdbo}) leads to
\be
   \lefteqn{
  (\tilde{V}_n \cdots \tilde{V}_{n+q-p-1} + h \, I ) \,
  \tilde{V}_{n+q-p} \cdots \tilde{V}_{n+q-1}
   } \hspace*{2cm} \nonumber \\
 &=& V_n \cdots V_{n+p-1} \, (V_{n+p} \cdots V_{n+q-1} + h \, I)
\ee
which can be mapped to (\ref{dtLV1}) via a redefinition
$T \mapsto T S^{p-q}$ (so that $\tilde{V}_n(t) = V_{n+p-q}(t+h)$)
and a subsequent replacement $(h,p,q) \mapsto (h^{-1},q,p)$.
Finally, if $p=-r<0$, equation (\ref{gdbo}) takes the form
\be
   \tilde{g}_n^{-1} \, \tilde{g}_{n+q} - g_n^{-1} \, g_{n+q}
 = h \, [ (g_{n-r}^{-1} \, g_n)^{-1} - (\tilde{g}_{n+q}^{-1} \,
   \tilde{g}_{n+q+r})^{-1}]
\ee
which in terms of $V_n$ reads
\be
  (I + h \tilde{V}_{n+q+r-1}^{-1} \cdots \tilde{V}_n^{-1})
  \tilde{V}_n \cdots \tilde{V}_{n+q-1}
  = V_n \cdots V_{n+q-1} (I+h V_{n+q-1}^{-1} \cdots V_{n-r}^{-1}) \,.
\ee
In the scalar case with $q=1$ this reduces to
\be
 \tilde{a}_n \, \Big(1 + h \prod_{i=0}^{r} \tilde{a}_{n+i}^{-1} \Big)
 = a_n \, \Big(1 + h \prod_{i=0}^{r} a_{n-i}^{-1}\Big)
\ee
and with a change of variable $v_n = a_{-n}$ we recover equation
(3.9) in \cite{Suri96}.
\vskip.1cm

\noindent
{\bf B.}
Let us now consider a dressing of $\dl$ (while leaving $\d$ unchanged):
\be
 \Dl z = (z - L \,T S^{q-p}z) \, \xi + h^{-1} \, (M \, T z-z) \, \eta
\ee
with $(Lz)_n = L_n \, z_n$ and $(Mz)_n = M_n \, z_n$. $({\cal M},\Dl,\d)$
is a bicomplex if and only if
\be
 L_n \, \tilde{M}_{n+q-p} = M_n \, \tilde{L}_n \, , \quad
 L_n - L_{n+p} = h^{-1} \, (M_{n+q} - M_n) \, .
     \label{tdbc2}
\ee
\vskip.1cm
\noindent
{\em Remark.}
Writing $L_n(t) = g_{n+p}(t-h) \, g_{n+q}(t)^{-1}$ and
$M_n(t) = g_{n+p}(t-h) \, \tilde{g}_{n+p}(t)^{-1}$,
the two bicomplexes $({\cal M},\dl,\D)$ and $({\cal M},\Dl,\d)$
are gauge equivalent.
\hfill $\blacksquare$
\vskip.1cm

The ansatz $L_n = G_n - G_{n+q}$, $M_n = I + h \, (G_{n+p}-G_n)$
solves (\ref{tdbc2}b) and yields the correct $h \to 0$ limit
for $\Dl$. Now (\ref{tdbc2}a) becomes
\be
 (G_{n+q}-G_n) \, [I + h \, (\tilde{G}_{n+q} - \tilde{G}_{n+q-p})]
 = [I + h \, (G_{n+p}-G_n)] \, (\tilde{G}_{n+q}-\tilde{G}_n) \, .
   \label{dG-eq}
\ee
Again, we may assume $q>0$. If $p>0$, in terms of $W_n = G_{n+1} - G_n$
(\ref{dG-eq}) reads
\be
     \Big( \sum_{i=0}^{q-1} W_{n+i} \Big) \,
     \Big( I + h \, \sum_{j=1-q}^{p-q} \tilde{W}_{n-j} \Big)
   = \Big( I + h \, \sum_{j=0}^{p-1} W_{n+j} \Big)
     \, \sum_{i=0}^{q-1} \tilde{W}_{n+i}  \, .
\ee
Even for $q=1$ and $p \geq 1$ and in the scalar case this
appears to be a new integrable time-discretization of
the corresponding LV equation (which is a special case
of (\ref{LV3})). It is different from the discretization
(3.1) in \cite{Suri96}.
If $p < 0$, (\ref{dG-eq}) is mapped via
$(p,q,h,t,G_n) \mapsto (q-p,q,-h,t+h,G^T_n)$
to the above case with positive $p$.

\section{LV equations as reductions of a three-dimensional equation}
Another trivial bicomplex $({\cal M}, \dl, \d)$ is obtained by replacing
(\ref{tdbc}) with
\be
\dl z = (z-T K \Lambda^{-1} z)\xi +h^{-1} (T z -z)\eta \, , \qquad
\d z  = (K z - z)\xi + (\Lambda z -z)\eta
\ee
where $(Tz)_{k,l}(t)=z_{k,l}(t+h)$, $(Kz)_{k,l}=z_{k+1,l}$ and
$(\Lambda z)_{k,l}=z_{k,l+1}$ are acting on the set ${\cal M}^0$
of $\mathbb{C}^m$-valued functions $z_{k,l}(t)$ depending on three
discrete variables $k,l \in \mathbb{Z}, \, t \in h \mathbb{Z}$.
The dressing $\D z = (Q K z -z) \, \xi + (P \Lambda z -z) \, \eta$
with $Q_{k,l} = g_{k,l}^{-1} \, g_{k+1,l}$ and
$P_{k,l} = g_{k,l}^{-1} \, g_{k,l+1}$ then leads to a bicomplex
$({\cal M},\dl,\D)$ associated with
\be
   \tilde{g}_{k,l}^{-1} \, \tilde{g}_{k+1,l} - g_{k,l}^{-1} \, g_{k+1,l}
 = h \, (g_{k,l}^{-1} \, g_{k,l+1} - \tilde{g}_{k+1,l-1}^{-1}
   \, \tilde{g}_{k+1,l}) \, .   \label{dhiro}
\ee
In the scalar case, this equation should be equivalent to the discrete
Hirota equation (cf \cite{DMH-BT}, section 4.2.5).
If all the fields only depend on $n := k q + l p$ with fixed
$p,q \in \mathbb{Z}$, we recover equation (\ref{gdbo}).
The continuous time limit of (\ref{dhiro}) is
\be
  (g_{k,l}^{-1} \, g_{k+1,l}) \, \mathbf{\dot{ }} \, =
   g_{k,l}^{-1} \, g_{k,l+1} - g_{k+1,l-1}^{-1} \, g_{k+1,l} \; .
\ee

\end{document}